\documentstyle[12pt,russian]{article}
\begin{document}
\centerline{\bf Noncommutative deformation and}
\centerline{\bf a topological nature of singularity Koiter}
\bigskip
 \centerline{\it Trinh V. K.}
 \centerline{\it Department of physics of polymer and crystal,
MGU}
\centerline{\it E-mail : khoa@polly.phys.msu.ru }
\large
1. From the point of view of phenomenology, till now there are
 many models of plastic deformation.
Almost in all monographs , for example [1-3,10], the
following models are considered: sliding,
regular, current, singular, analytical and deformation
plasticity.
The determinant ratio in deformation are a major
problem, which was
 submitted for discussion. The its most simple kind was
 presented in the form of Genki-Nadai:
$$
\sigma_{ij} =2G_s(T) e_{ij} \qquad\mbox {at}~~ T =
\bar T \eqno (1)
$$
$$
d\sigma_{ij} =2Ge_{ij} \qquad\mbox {at}~~ T =\ne\bar T
$$
Where $d\sigma_{ij}, ~e_{ij} $ - deviatory of a
stress and deformation, $G_s $- current
 module  of pure shift,
$G $- the elastic module of shift, $T $- intensity
 tangent   stress, $ \bar T $-
the maximal value Т for a history of loading  pre-eminent
some value $T_0 $. In singularity model it looks like
$$
de_{ij} = \sum\limits_{n=1}^{N}df_n
\frac{\partial f_n}{\partial\sigma_{ij}} \eqno (2)
$$
Where $f_n $- the current limiting surface. Here not
zero are those $df_n $, for
which on the appropriate surface $f_n $ be performed
loading, i.e.
$$
( \frac{\partial f_n}{\partial\sigma_{ij}}) d\sigma_{ij}
\ge 0
$$
In model Iliusin it has the kind
$$
\sigma_{ij} = \sum\limits_{n}A_n\frac{d^n e_{ij}}{de^n}:
 ~e =\int\sqrt{de_{\alpha}de_{\alpha}} \eqno(3)
$$
However, if we use concept of potential of deformation as
in [1-3,10-11], on
 definition of function $H $, we shall receive following
determining ratio
$$
\sigma_{ij} = \frac{\partial H}{\partial e_{ij}} \eqno(4)
$$
In our model [4], in which we consider process
of plastic deformation
as process of phase transition, potential of deformation
(or the free energy) plays
role of function Hamilton or action of system. So, if
 we admit Hamilton as
$H =\alpha e^2 +\beta e^4 +\ldots, $ где $ \alpha, \beta $-
factors of elasticity, е- deformation tensor,
that as a first approximation we shall receive the formulas
(1) (2) (3) (if $H $ is expressed through the metrics of
spaces). Roughly speaking, form (4) is general
determining ratio for
process of plastic deformation. Here there are two problems:
how the fields
 $ \sigma, e $ in the time and in space is distributed ?
  What nature of
the singularity Koiter ?
We try to solve these problems consistently.
 For this purpose we should construct more general
model of deformation.

2. As is known, all experimental data have shown, that in
a plastic status
the stress and deformation is strongly fluctuation.
 Besides the diagram $ \sigma=f(e) $
for pressure - deformation is similar to the diagram $P=f(V) $
for pressure - volume for a liquid [5].
So, it is possible to consider a  process of deformation
as process of phase transition, or more
 generalized speaking, it is process of formation
 of structure [4]. Density of distribution of probability
transition to plasticity is solution of the equation
 Fokker-Plank. However, if
using potential character of process of deformation it is
possible instead of the equation
Fokker-Plank to apply the equation Schrodinger.
In this approach tensor of  deformation is chosen in
the roles of parameter of the order. In work [1] A.A. Iliusin
 has presented the concept of trajectory of deformation in
 space $E_5 $.
 As a matter of fact the space
$E_5$ is the one fiber in the deformation bundle [4]
 Further, because of fluctuation of
the statuses of deformation, it is necessary for me to used "
 the secondary calculation " to receive a field
deformations in a plastic status from the elasticity.
 Nevertheless it is necessary to notice,
that a nature of process of plastic deformation is
 noncommutative nature.

So, the process of deformation $P = \{e_{ij}(x, t)\} $
is given. That is noncommutative deformation space.
The deformation wave $E_{ij}(x, t) $ is noncommutative wave.
For complete
 descriptions of distribution of this wave we have to
 use the space being tensor
  product $C(R_4) \otimes M_n $, and $C(R_4) $ - algebra
 of smooth function determined
on usual space-time, and $M_n $ - algebra $n\times n $ of
 matrixes. In this general
space the matrix function $E(x, t) $ looks like (for 2-dimention
  case):
$
E(x, t) = \left(\begin{array}{cccc}
        a&b\\
        c&d
        \end{array}
         \right)
\in C(R_4) \otimes M_2 $.
Now our task is construction of noncommutative deformation bundle
 on the
$C(R_{4}) \otimes M_2 $. This problem was performed in
 frameworks of
 noncommutative geometry
Connes [6]. There are two important objects in this geometry.
It is associative algebra
$ {\cal A} $ and universal algebra of the differential forms
 $ \Omega^1_D ({\cal A}) $ on
the algebra $ {\cal A} $. Here space of deformation
$P=C(R_4) \otimes M_n $ plays
role of associative algebra $ {\cal A} $, at the same time
universal algebra
$ \Omega^1_D({\cal A}) $ looks like [7]:
$$
\Omega^1_D({\cal A}) = \Omega^1_D(C(R_4)) \otimes
\Omega^1_D(M_n)
$$
It is tensor product of algebra of the differential forms
 above algebra of smooth function
and algebra of the differential forms above algebra
$n\times n $ of matrixes. In turn
this algebra $ \Omega^1_D({\cal A}) $ can be present
 through the direct sum of
a horizontal and vertical of  part
$$
\Omega^1_D({\cal A}) = \Omega^1_H\oplus \Omega^1_V
$$
and
$$
\Omega^1_H =\Omega^1_D(C (R_4)) \otimes M_n; ~
\Omega^1_V=C(R_4) \otimes\Omega^1_D(M_n)
$$
Thus, differential calculation on algebra $ {\cal A} $,
 being in usual external differential,
is chosen. More precisely,  if through
$Der({\cal A}) $ we had designated Lie algebra of derivative
on $ {\cal A} $, we shall receive
$$
Der({\cal A}) = (Der(C(R_4)) \otimes 1) \oplus (C (R_4)
\otimes Der(M_n))
$$
Differential $df $ of an element $f\in {\cal A} $ also
is divided for the sum
$$
df=d_H f+d_V f
$$
where $d_h f $ and $d_V f $ belong to $ \Omega^1_H $ and
$ \Omega^1_V $  respectively. If
$E_k, k\in \{1,2, \ldots, n^2-1 \} $ is basis of algebra
$M_n $, then $ \partial_k=ad(iE_k) $
 is a basis of algebra $Der(M_n) $ and
$ [\partial_k, \partial_l] = \sum C_{klm}\partial_m $.
 So,
 the basis $ \theta^k $ of algebras $ \Omega^1_D(M_n)
\subset\Omega^1_D({\cal A}) $ is determined
as $ \theta^k(\partial_l) =\delta^k_l 1 $. If through
$ \theta^i = (\theta^{\alpha}, \theta^a) $,
where $ \alpha=0,1,2,3; a\in \{1,2, \dots, n^2-1 \} $,
the basis of algebra $ \Omega^1_D({\cal A}) $ is designated,
then the basis in $Der({\cal A}) $ will be as $e_i =
(e_{\alpha}, e_a) $ and $e_{\alpha} =e^{\mu}_
{\alpha} \partial_{\mu} $
 are 4-generators for algebra $C(R_4) $, and
 $e_{\alpha} =ad(E_{\alpha}) $  is a basis of
$Der(M_n) $. In this case $ \theta^i $ accepts a kind
$$
\theta^{\alpha} = \theta^{\alpha}_{\mu} dx^{-\mu},
~ \theta^{a} =E_b E^a dE^b
$$
Differential $d_h $ and $d_V $ will accept a kind
$$
d_H f=e_{\alpha}f\theta^{\alpha}, ~d_V f=
e_a f\theta^a, ~ f\in {\cal A}
$$
Besides there is an canonical element $ \theta\in
\Omega^1_D(M-n) $ determined as
$ \theta=E_k\theta^k $. Within the framework of geometry
Connes, noncommutative bundle is right,
 or left $ {\cal A} $ - module, or is more exact,
 Hermit $ {\cal A} $ - module.
On this module  following connection
is defined [7]
$$
\omega=A +\chi
$$
and $A $- it is an element of $ \Omega^1_H $ and $ \chi $-
an element of $ \Omega^1_V $. In  turn,
the element $ \chi $ is divided for the sum
$$
\chi =\theta +\phi
$$
because of the fact, that $ \Omega^1_D(M_n) \subseteq
\Omega^1_V $. Here $ \phi $ is a field
Higgs. If through $ {\cal U} $ it displays the group of gauge maps
, then an element
$ \omega, A, \theta, \phi $ will be transformed under action
$g\in {\cal U} $ by a rule
$$
\omega' =g^{-1} \omega g+g^{-1} dg \eqno(5)
$$
$$
A' =g^{-1} +g^{-1}d_H g, ~ \theta' = \theta,
~ \phi' =g^{1-}\phi g
$$
The intensity of a field of deformation is determined
through the 2-form $ \Omega $
$$
\Omega=d\omega +\omega^2
$$
Or
$$
\Omega=F+D_H\phi +\Omega_V \eqno(6)
$$
And
$$
F=d_H A + [A, A] = \frac{1}{2}F_{\alpha\beta}
\theta^{\alpha}\theta^{\beta}
$$
$$
\Omega_V =\frac{1}{2}\Omega_{ab}\theta^a\wedge\theta^b,
~ \Omega_{ab} = [\phi_a, \phi_b] -C^c_{ab}\phi_c
$$
$$
D_H\phi=d_h\phi+A\phi +\phi A =\theta^{\alpha}(e_{\alpha} +
 [A_{\alpha}, \phi])
$$
$$
\phi =\phi_{\alpha}\theta^{\alpha}, ~A=
A_{\alpha}\theta^{\alpha}
$$
Action of a field will be as
$$
L = \frac{1}{2}(\Omega_{ij}, \Omega_{ij}) \eqno(7)
$$
Thus, in frameworks model  [4], we shall receive probability
of transition to plasticity
$$
Z = \int\exp(-kL)
$$
Where $k $- it is factor of elasticity.

3. We discuss the problem put above. For simplicity we
 shall be limited to consideration
only pure deformation also we shall not take into
 account the interaction with a field Higgs. Actually,
 all above mentioned models were considered
only in such similar
situations. In these models was assumed about existence of
limiting surfaces
deformations (or loading). It is equivalent to the
requirement of finiteness of action of
deformation fields . Then, for search of a configuration
of a field of deformation, it is necessary
to solve the equation, which is a condition
extreme of action $L $,
$$
\delta L[A_{\mu}] =0 \eqno(9)
$$
However, we have also other condition for action
$$
-\int d^4 x Tr [(F_{\mu\nu} + \pm {\tilde F}_{\mu\nu})^2]
$$
where $ {\tilde F_{\mu\nu}} $ is dual with $F_{\mu\nu} $.
Because of that
$Tr[F_{\mu\nu}, F_{\mu\nu}] =Tr[\tilde F_{\mu\nu},
\tilde F_{\mu\nu}] $, we receive
$$
-\int d^4 x Tr[F_{\mu\nu}, F_{\mu\nu}]
\ge \mp\int d^4 x Tr[\tilde F_{\mu\nu}, \tilde F_{\mu\nu}]
$$
Is like as in work Belavin etc. [9], we are interested in
importance of
 self-dual and
self-antidual  configuration of a field of deformation.
Thus, one of a extreme condition of
action shall write down as
$$
\tilde F_{\mu\nu} = \mp F_{\mu\nu} \eqno(10)
$$
Belavin etc. used this condition for giving  the
instanton
for gauge field. So, instead of the equation
(9) we shall  solve  more simple
the equation (10). For it we use method in [8] for search
 of multiinstanton  in
special case, when $n, r=1 $. Let the field of deformation
 $A=A_{-mu} dx^{-mu} \in
\Omega^1_D(R_4) \otimes 1 $, has a individual kind
$$
A_{\mu}(x) =iB_{\mu\nu}\partial_{\nu}(\ln\phi (x))
\eqno(11)
$$
where $ \phi(x) $ - it is a scalar function, and the matrix
 $B_{\mu\nu} $ looks like
$$
B_{\mu\nu}=\frac{1}{2}\left\{
                            \begin{array}{cccc}
                             0 & +\sigma_3 & -\sigma_2 & -\sigma_1\\
                             -\sigma_3 & 0 & +\sigma_1 & -\sigma_2\\
                             +\sigma_2 & -\sigma_1 & 0 & -\sigma_3 \\
                             +\sigma_1 & +\sigma_2 & +\sigma_3 & 0
                                    \end{array}
                      \right\}
$$
where $ \sigma_1, \sigma_2, \sigma_3 $- it is matrixes Pauli.
 The gauge group is admitted
as $SU(2) $. Setting  $A_{\mu} $ in expression (6)
that condition of self-dual (10) will be as
$$
\partial_{\sigma}\partial_{\sigma}(\ln\phi) +
(\partial_{\sigma}\ln\phi)^2=0
$$
or
$$
\frac{{\nabla}^2\phi}{\phi} =0,
~ {\mbox {where}}{\nabla}^2 =\partial_{\sigma}
\partial_{\sigma} \eqno(12)
$$
Using a method in [9] we receive instanton, being the
solution  (12)
$$
\phi(x) =1 +\frac{{\lambda_1}^2}{{\mid x\mid}^2} \eqno(13)
$$
and multiinstanton
$$
\phi(x) =1 +\sum\limits_{i=1}^{N}
\frac{{\lambda_i}^2}{{\mid x_{\mu} -a_{i\mu} \mid}^2}
\eqno(14)
$$
where $a_{i\mu}, \lambda_i $- it is any valid constant.
 Setting  $ \phi(x) $
from (12) and (13) in (11) we receive deformation wave
as  matrixes:

One-instanton
$$
F_{\mu}(x) = -2i\lambda^2 B_{\mu\nu}
\frac{y_{\nu}}{y^2(y^2 +\lambda^2)}
$$

Multi-instanton
$$
A_{\mu}(x) = -2iB_{\mu\nu}(\sum\limits_i^N
\frac{\lambda_I^2 y_{i\nu}}{{y_i}^4}) /
( 1 +\sum\limits_j^N\frac{\lambda_i^2}{{\mid y_j\mid}^2})
$$
where $ (y_i)_{\mu}\equiv (x-a_i)_{\mu},
~i, j=1,2, \ldots, N $

4. What real nature of sectors Koiter on a limiting surface?.
With the account of solution of instanton
, this problem can be easily clear. Really, using these
instanton, we receive topological number Pontriagin
$$
Q = -\frac{1}{16\pi^2} \int d^4 x Tr[F_{\mu\nu} F_{\mu\nu}]
$$
This number tell me, that space of deformation during
plastic deformation
will be is divided on $Q $ of independent sectors.
Thus, limiting surface,
constructed in the same space, too is divided on
$Q $ of parts. It is sectors Koiter,
described in works [1-3,10-11]. We see, that occurrence
of singularity Koiter
has the topological reasons. Now we can speak, that process
 of plastic deformation
is a process of birth and distribution of solitons and
like-solitons.  It is necessary to notice, that
the defect,  from the point of
view of fluctuation, is one of kinds of solitons.
The soliton wave  is considered as the concentrated
located energy, which
is distributed in space without dissipation  with constant
velocity. Thus,
the development of process of plastic deformation occurs
with constant velocity and it needlessly
 increases a load. It will allow to understand, why a part
 of plasticity of the diagram
stress - deformation has the horizontal form.

\end{document}